\documentclass[11pt,oneside,fleqn]{article}

\usepackage[utf8]{inputenc}
\usepackage{hyperref}
\usepackage{amsmath,amssymb}
\usepackage{graphicx}
\usepackage{color}

\allowdisplaybreaks

\hypersetup{colorlinks, linkcolor=blue, citecolor=blue, urlcolor=blue}

\newcommand{\dd}[2]{\frac{\mathrm{d} #1}{\mathrm{d} #2}}
\newcommand{\pdl}[2]{\frac{\partial #1}{\partial #2}}
\newcommand{\pds}[2]{\partial #1 / \partial #2}
\newcommand{\ddl}[2]{\frac{\delta #1}{\delta #2}}
\newcommand{\ddd}{\mathrm{d}}
\newcommand{\p}{\partial}

\newcommand{\const}{\mathop{\rm const}\nolimits}
\newcommand{\todo}[1][\null]{\ensuremath{\clubsuit}}

\newcommand{\nn}{\mathbf{\nabla}}

\newcommand{\ve}{\varepsilon}

\newcommand{\ZZ}{\mathcal{Z}}

\newcommand{\XX}{\mathcal{X}}
\newcommand{\DDD}{\mathcal{D}}

\begin{document}

\par\noindent {\LARGE\bf
Minimal atmospheric finite-mode models preserving \\ symmetry and generalized Hamiltonian structures
\par}
{\vspace{4mm}\par\noindent {\bf Alexander Bihlo~$^\dag$ and Johannes Staufer~$^\dag\, ^\ddag$
} \par\vspace{2mm}\par}

{\vspace{2mm}\par\noindent {\it
$^{\dag}$~Faculty of Mathematics, University of Vienna, Nordbergstra{\ss}e 15, A-1090 Vienna, Austria\\
}}
{\noindent \vspace{2mm}{\it
$\phantom{^\dag}$~\textup{E-mail}: alexander.bihlo@univie.ac.at
}\par}

{\vspace{2mm}\par\noindent {\it
$^\ddag$~Now at: Institute for Atmospheric and Climate Science, ETH Z\"{u}rich, Universit\"{a}tstrasse 16, CH-8092 Zurich, Switzerland\\
}}
{\noindent \vspace{2mm}{\it
$\phantom{^\dag}$~\textup{E-mail}: johannes.staufer@env.ethz.ch
}\par}

\vspace{2mm}\par\noindent\hspace*{8mm}\parbox{140mm}{\small\looseness=-1
A typical problem with the conventional Galerkin approach for the construction of finite-mode models is to keep structural properties unaffected in the process of discretization. We present two examples of finite-mode approximations that in some respect preserve the geometric attributes inherited from their continuous models: a three-component model of the barotropic vorticity equation known as Lorenz' maximum simplification equations [Tellus, \textbf{12}, 243--254 (1960)] and a six-component model of the two-dimensional Rayleigh--B\'{e}nard convection problem. It is reviewed that the Lorenz--1960 model respects both the maximal set of admitted point symmetries and an extension of the noncanonical Hamiltonian form (Nambu form). In a similar fashion, it is proved that the famous Lorenz--1963 model violates the structural properties of the Saltzman equations and hence cannot be considered as the maximum simplification of the Rayleigh--B\'{e}nard convection problem. Using a six-component truncation, we show that it is again possible retaining both symmetries and the Nambu representation in the course of discretization. The conservative part of this six-component reduction is related to the Lagrange top equations. Dissipation is incorporated using a metric tensor.
}\par\vspace{2mm}

\section{Introduction}\label{sec:introduction}

Various models of the atmospheric sciences are based on nonlinear partial differential equations. Besides numerical simulations of such models, it has been tried over the past fifty years to capture at least some of their characteristic features by deriving reduced and much simplified systems of equations. A common way for deriving such reduced models is based on the Galerkin approach: One expands the dynamic variables of a model in a truncated Fourier (or some other) series, substitutes this expansion into the governing equations and studies the dynamics of the corresponding system of ordinary differential equations for the expansion coefficients. Although the number of expansion coefficients is usually minimal to allow for an analytic investigation, these reduced models have been used in order to explain some common properties of atmospheric models.

To the best of our knowledge there is up to now no universal criterion for the selection of modes or the choice of truncation of the series expansion. However, at least some cornerstones for the Galerkin approach are already settled. It is desirable for finite-mode models to retain structural properties of the original set of equations, from which they are derived~\cite{gluh06Ay,gluh02Ay}. Such properties are, e.g., quadratic nonlinearities, conservation of energy and one or more vorticity quantities in the nondissipative limit and preservation of the Hamiltonian form.

Recently, an extension of the Hamiltonian structure based on the idea of Nambu \cite{namb73Ay} to incorporate multiple conserved quantities in a system representation also came into focus. It was shown in \cite{bihl08Ay,nevi98Ay,nevi93Ay,nevi09Ay,sala10Ay,somm09Ay} that various equations of ideal hydrodynamics and magneto-hydrodynamics allow for a formal Nambu representation. It therefore seems reasonable to derive finite-mode models that also retain this structure. Moreover, almost all models in the atmospheric sciences possess symmetry properties. These symmetries should thus be taken into account in low-dimensional modeling too, which is an issue in the field of equivariant dynamical systems (see, e.g.,~\cite{golu00Ay}).

The general motivation for this work is that low-order models are still in widespread use in the atmospheric sciences. It has been mentioned above that their original purpose was to identify characteristic features of the atmospheric flow in the pre-supercomputer era. While the advent of supercomputers partially renders this aim obsolete, finite-mode models are still valuable for testing advanced methods in the atmospheric sciences, related to issues of predictability, ensemble prediction, data assimilation or stochastic parameterization~\cite{ande99Ay,mill94Ay,palm93Ay,palm01Ay}. Such finite-mode models offer the possibility for a conceptual understanding of techniques that are to be used in comprehensive atmospheric numerical models later on. For such testing issues, in turn, it is essential to have finite-mode models that preserve the structure of the underlying set of partial differential equations at least in some minimal way.

In this paper we give two examples of finite-mode models that retain the above mentioned features of their parent model: The first is the three-component Lorenz--1960 model, derived as the maximum simplification of the vorticity equation \cite{lore60Ay}. The second is a six-component extension of the Lorenz--1963 model~\cite{lore63Ay}. The authors are aware that there exists a great variety of other finite-mode (Lorenz) models, such as e.g.~\cite{lore80Ay,lore86Ay,lore87Ay}, possessing richer geometric structure and allowing to address other important issues in the atmospheric sciences, such as the existence of a slow manifold, atmospheric attractors, balanced dynamics and the initialization problem of numerical weather prediction. Results in these directions can be found, besides in the original papers by Lorenz, e.g., in~\cite{bokh96Ay,cama95Ay,vann04Ay,wiro00Ay}. The choice to investigate the Lorenz--1960 and Lorenz--1963 models, however, is reasonable since the latter still is one of the most prominent finite-mode models used in dynamic meteorology for testing issues as reviewed above. As we are going to show, the Lorenz--1963 model in various respects does not constitute a sound geometric model, the derivation of a revised version of this system appears to be well justified. The Lorenz--1960 model, on the other hand has been chosen as it is the simplest system for which the techniques to be applied in this paper can be demonstrated.

The Lorenz--1963 model is a dissipative model and as such it necessarily violates conservative properties. On the other hand this is a rather typical situation for more comprehensive atmospheric numerical models too. Usually, the conservative dynamical core of such models is coupled to a number of dissipative processes such as friction, precipitation and radiation. Nonetheless, it is a necessary condition that the numerics for the dynamical core itself do not violate the structural properties of the underlying conservative dynamics~\cite{zeit05Ay}. Any valuable toy model of the atmosphere should reflect this, e.g.\ by consisting of the superposition of a conservative part and a dissipative part. This is one of the guiding principles for our derivation of the generalized Lorenz--1963 model.

The organization of the paper is as follows: Properties of discrete and continuous Nambu mechanics are briefly reviewed in section~\ref{sec:nambu}. Section~\ref{sec:vorticity} includes a description of the Lorenz--1960 model, establishing its Nambu structure and its compatibility with the admitted point symmetries of the barotropic vorticity equation. In section~\ref{sec:RBC}, it is shown that the Lorenz--1963 model is neither compatible with the corresponding Nambu (Hamilton) form of the Saltzman convection equations nor with its point symmetries. We hereafter identify the maximum simplification of the Saltzman convection equations~\cite{salt62Ay} that reflects both symmetries and the proper Nambu structure of the continuous model. Finally, in section~\ref{sec:conclusion} we sum up our results and discuss some open questions.

\section{Nambu mechanics}\label{sec:nambu}

Since Nambu mechanics emerged from discrete Hamiltonian mechanics, it is convenient to start with a short description of the latter. The evolution equation of a general $n$-dimensional Hamiltonian system is given by
\[
	\dd{F}{t}=\left\{F,H\right\},
\]
where $F=F(z_i)$ is an arbitrary function of the phase space variables $z_i$, $i=1,\dots, n$, $H$ is the Hamiltonian function and $\left\{.,.\right\}$ is a Poisson bracket, which satisfies bilinearity, skew-symmetry and the Jacobi identity. For discrete Hamiltonian systems, the Poisson bracket is characterized by an antisymmetric rank two tensor that can depend on the coordinates of the underlying phase space. In modern Hamiltonian dynamics, this tensor is allowed to be singular, leading to the notion of a \textit{Casimir} function $C$, which Poisson-commutes with all arbitrary functions $G(z_{i})$
\[
	\left\{C,G\right\}=0, \quad \forall\ G.
\]
Setting $G=H$, it follows that every Casimir is in particular also a conserved quantity.

Guided by Liouville's theorem stating volume-preservation in phase space, Nambu \cite{namb73Ay} proposed a formalism for discrete mechanical systems allowing \textit{multiple} conserved quantities to determine, at the same level of significance, the evolution of a dynamical system. More precisely, let us consider a point mechanical system with $n$ degrees of freedom and $n-1$ functionally independent conserved quantities $H_j, j=1,\dots, n-1$. The evolution equation for an arbitrary function $F$ according to Nambu is
\[
	\dd{F}{t}=\frac{\partial(F,H_{1},H_{2},\dots,H_{n-1})}{\partial({z_{1},z_{2},\dots,z_{n}})}=:\left\{F,H_{1},H_{2}, \dots,H_{n-1}\right\}.
\]
The above bracket operation is called \textit{Nambu bracket}, which due to the properties of the Jacobian is non-singular, multi-linear and totally antisymmetric. It was demonstrated in~\cite{takt94Ay}, that a Nambu bracket also fulfills a generalization of the Jacobi identity, which reads
\begin{align}\label{eq:genjac}
\begin{split}
	\{\{F_1,\dots,F_{n-1},F_{n}\},F_{n+1},\dots,F_{2n-1}\}+\{F_n,\{F_1\dots,F_{n-1},F_{n+1}\},F_{n+2},\dots,F_{2n-1}\} \\+ \cdots +\{F_n,\dots,F_{2n-2},\{F_{1},\dots, F_{n-1},F_{2n-1}\}\} = \{F_1,\dots,F_{n-1},\{F_{n},\dots,F_{2n-1}\}\}
\end{split}
\end{align}
for any set of $2n-1$ functions $F_i$. Various discrete models that allow for a Nambu formulation were identified, e.g.\ the free rigid body~\cite{namb73Ay}, a system of three point vortices~\cite{nevi98Ay}, and the conservative Lorenz--1963 model \cite{nevi94Ay}, which is discussed in some detail below.

It appears that the application of ideas of discrete Nambu mechanics to field equations was first considered in~\cite{bial91Ay} (and even earlier in a talk~\cite{meis84Ay}), and later independently by N\'{e}vir and Blender \cite{nevi93Ay}. It was noted that the singularity of many continuous Poisson brackets of fluid mechanics may be formally removed by extending them to tribrackets using explicitly \emph{one} of their Casimir functionals as additional conserved quantity. That is, despite the fact that partial differential equations represent systems with infinitely many degrees of freedom, up to now there only exist models using one additional conserved quantity. This way, the term \textit{continuous Nambu mechanics} (referring to a Nambu representation of field equations) is at once misleading, though it is already used in several papers.

The restriction to tribrackets may be traced back to the underlying Lie algebras on which the Poisson brackets in Eulerian variables are based on~\cite{mars99Ay,morr98Ay}. Hence, the fixed relation between the dimension of the phase space and the number of conserved quantities used for a system representation is lost in continuous Nambu mechanics. In the atmospheric sciences, this generalization is called \emph{energy-vorticity theory}, as the employed Casimir functional is frequently related to some vortex integral. Since in the atmospheric sciences the evolution of the rotational wind field is dominant over different scales, the energy-vorticity description may be well suited for a better understanding of e.g.\ turbulence. Among others, models that can be cast into energy-vorticity form include the inviscid non-divergent 2d and 3d barotropic vorticity equations, the quasi-geostrophic potential vorticity equation and the governing equations of ideal fluid mechanics as well as equations of magnetohydrodynamics~\cite{bihl08Ay,nevi98Ay,nevi04Ay,nevi09Ay,sala10Ay}.

The main problem with continuous Nambu mechanics is that it is up to now not clear whether it is possible to state an appropriate condition analog to the generalized Jacobi identity~\eqref{eq:genjac} of discrete Nambu mechanics. While this is obviously a serious point assessing the self-reliance of continuous Nambu mechanics compared to usual noncanonical Hamiltonian field theory, for the application of the Nambu formalism this point is not of prior importance. Indeed, the main benefit of a continuous Nambu formulation so far lies in the possibility of the construction of conservative numerical schemes. Namely, numerically conserving the antisymmetric Nambu bracket automatically leads to a numerical conservation of energy and the second constitutive conserved quantity of the bracket. This allows to explain the construction of the celebrated Arakawa discretization~\cite{arak66Ay} of the Jacobian operator and enables to generalize the Arakawa method in a systematic way to other models possessing a Nambu representation~\cite{gass08Ay,salm05Ay,salm07Ay,somm09Ay}. That is, although the Nambu form might appear to be an algebraic curiosity from the theoretical point of view, it is nevertheless of high value in the numerical application. This is also the main reason, why we aim to care about this form in the course of the present paper. The generalized Lorenz--1963 model derived in section~\ref{sec:RBC} is of Nambu form and hence it automatically conserves certain representatives of both classes of Casimir functionals of the Rayleigh-B\'{e}nard convection equations.

It could be argued, that the numerical preservation of only one additional conserved quantity is only a little success in view of the infinite number of conserved quantities of two-dimensional ideal hydrodynamics. As was demonstrated in~\cite{abra03Ay}, not only energy, circulation and enstrophy are statistically relevant for the large scale behavior of ideal fluid mechanics, but also is the third integrated power of the vorticity. Though this objection can hardly be rebutted, the authors are not aware of any truncation conserving an appropriate number of conserved quantities besides the method proposed by Zeitlin~\cite{zeit91Ay,zeit05Ay}, which however can neither be adopted for all models in fluid mechanics nor in arbitrary geometries. Although similar objections also hold against the Nambu bracket approach, the Nambu discretization method nevertheless might be considered as an enrichment of existing numerical methods in fluid mechanics (see~\cite{dubi07Ay,somm10Ay} for a discussion of the Nambu discretization in relation to the statistics). Moreover, the Nambu bracket approach goes beyond various numerical methods for Hamiltonian field equations, in which solely conservation of energy can be assured.

\section{Structural properties of the vorticity equation}\label{sec:vorticity}

The inviscid barotropic vorticity equation on the \textit{f}-plane for an incompressible flow can be written in form of a conservation law
\begin{equation}\label{eq:BVE}
	\frac{\partial\zeta}{\partial t} = - \left[\psi, \zeta \right],
\end{equation}
where $\psi(t,x,y)$ is the stream function generating two-dimensional nondivergent flow, $\zeta=\nabla^2 \psi$ is the vorticity and $\left[a,b\right]=\pds{a}{x} \hspace{3pt} \pds{b}{y} - \pds{a}{y} \hspace{3pt} \pds{b}{x}$ denotes the Jacobian.

\subsection{The Nambu structure}\label{sec:TheContinuousEquation}

Eqn.~\eqref{eq:BVE} possesses an infinite number of conserved quantities, i.e.\ kinetic energy and all moments of vorticity are preserved (see, e.g.,~\cite{arno99Ay}). In \cite{nevi93Ay}, the Nambu (or energy-vorticity) bracket
\begin{equation}	\label{eq:vortbrack}
	\left\{\mathcal{F}_1,\mathcal{F}_2,\mathcal{F}_3\right\}:=-\int_{\Omega}\frac{\delta\mathcal{F}_1}{\delta \zeta} \left[\frac{\delta \mathcal{F}_2}{\delta\zeta},\frac{\delta \mathcal{F}_3}{\delta \zeta}\right]\ddd f,
\end{equation}
was introduced for arbitrary functionals $\mathcal{F}_i[\zeta], i=1,\dots,3$. In the above equation, $\delta/\delta \zeta$ denotes the usual variational derivative, $\mathrm{d}f=\mathrm{d}x\mathrm{d}y$ is the area element to be integrated within the 2D-domain $\Omega$. Using appropriate boundary conditions (e.g.\ cyclic), it can be shown that the above bracket is totally antisymmetric. Geometrically, this Nambu bracket is essentially a reformulation of the singular Lie--Poisson bracket of ideal fluid mechanics, which is based on the infinite-dimensional Lie algebra associated to the group of area preserving diffeomorphisms on~$\Omega$.

Using the bracket~\eqref{eq:vortbrack} it is possible to reformulate eqn.~\eqref{eq:BVE} as
\begin{equation}
	\frac{\partial \zeta}{\partial t} = \left\{\mathcal{\zeta},\mathcal{E},\mathcal{H}\right\}.
\end{equation}
In the above equation, $\mathcal{H}$ and $\mathcal{E}$ denote the global conserved quantities energy and the second moment of vorticity (enstrophy), respectively, which are given by
\begin{equation*}
	\mathcal{H}[\zeta]=\frac{1}{2}\int_{\Omega}(\nabla \psi)^2\mathrm{d}f, \qquad \mathcal{E}[\zeta]=\frac{1}{2}\int_{\Omega}\zeta^2\mathrm{d}f.
\end{equation*}

\subsection{Maximum simplification}\label{sec:SpectralFormOfTheEnergyVorticityBracket}

One pioneering work in the field of finite-mode approximations was done by Lorenz~\cite{lore60Ay}, who introduced a minimal system of hydrodynamic equations based on~\eqref{eq:BVE}. Using a severe truncation of the Fourier series expansion of $\zeta$, the following set of ordinary differential equations for the remaining three modes $A,F,G$ was derived:
\begin{align}\label{eq:Lorenz60}
\begin{split}
	&\dd{A}{t}=\left(\frac{1}{k^2+l^2}-\frac{1}{k^2}\right)klFG, \\
	&\dd{F}{t}=\left(\frac{1}{l^2}-\frac{1}{k^2+l^2}\right)klAG, \\
	&\dd{G}{t}=\frac{1}{2}\left(\frac{1}{k^2}-\frac{1}{l^2}\right)klAF,
\end{split}
\end{align}
where $k, l$ are constant wave numbers. We now show, that this model preserves in some sense the structure of its continuous counterpart, as the above equations can be derived directly from the spectral Nambu bracket of the barotropic vorticity equation. For this purpose, we expand $\zeta$ in a double Fourier series on the torus
\begin{equation*}
	\zeta(c)=\sum_{\textbf{M}}c_{\textbf{M}}e^{i\textbf{M} \cdot
\textbf{x}},
\end{equation*}
where $\textbf{M}=(m_{1}k,m_{2}l)^{\mathrm{T}}$ and $\textbf{x}=(x,y)^{\mathrm{T}}$ are the wavenumber and position vector, respectively. Moreover, $c_{\textbf{M}}=c_{\textbf{-M}}^\dag= \frac{1}{2}\left(A_{m_{1},m_{2}}-iB_{m_{1} ,m_{2}}\right)$, where $\dag$ denotes the complex conjugate. The variational derivative may be expanded as
\begin{equation*}
	\frac{\delta \mathcal{F}[\zeta]}{\delta \zeta(c)}=\frac1\Omega\sum_{\textbf{M}}\frac{\partial \mathcal F}{\partial c_{\textbf{M}}}e^{-i\textbf{M} \cdot \textbf{x}},
\end{equation*}
where in the right hand side we consider $\mathcal F$ as a smooth function of $c_\mathbf{M}$. Plugging these expressions into eqn.~\eqref{eq:vortbrack} we find the spectral form of the energy--vorticity bracket
\begin{equation} \label{eq:sp60}
	\left\{\mathcal{F}_1,\mathcal{F}_2,\mathcal{F}_3\right\}=\sum_{\textbf{K,M}}kl(m_{1}k_{2}-m_{2}k_{1})\pdl{\mathcal F_1}{c_{\textbf{K}}}\pdl{\mathcal F_2}{c_{-(\textbf{M}+\textbf{K})}}\pdl{\mathcal F_3}{c_{\textbf{M}}}.
\end{equation}
To simplify the notation, we have assumed that $\Omega$ is the unit square. Enstrophy and energy in their spectral representations are given by $\mathcal E=1/2 \sum_{\textbf{K}}c_{\textbf{K}}c_{-\textbf{K}}$ and $\mathcal H=1/2\sum_{\textbf{K}}\frac{1}{\textbf{K}^2}c_{\textbf{K}}c_{-\textbf{K}}$, respectively. To obtain eqn.~\eqref{eq:Lorenz60} from bracket~\eqref{eq:sp60}, we have to truncate the conserved quantities $\mathcal E$ and $\mathcal H$ on the set of indices $k_{1}, k_{2} \in \left\{-1,0,1\right\}$ under the following restrictions introduced by Lorenz: $(i)$ If the coefficients are real at the onset of evolution they remain real for all times. $(ii)$ If $c_{1,1} = -c_{1,-1}$ at the onset of evolution this relation holds true for all times. Introducing the new variables $A=\mathrm{Re}(c_{0,1})$, $F=\mathrm{Re}(c_{1,0})$ and $G=\mathrm{Re}(c_{1,-1})$, it is straightforward to recover eqn.~\eqref{eq:Lorenz60} from eqn.~\eqref{eq:sp60}. The maximum simplification of the vorticity equation in Nambu form then reads
\begin{equation*}
	\frac{\mathrm{d}\textbf{z}}{\mathrm{d}t}=kl (\nabla_{\mathbf z} \mathcal E \times \nabla_{\mathbf z} \mathcal H) =:\left\{\textbf{z},\mathcal E,\mathcal H\right\},
\end{equation*}
with $\textbf{z}=(A,F,G)^\mathrm{T}$, where
\[
	\mathcal E =\frac12(A^2+F^2+2G^2), \quad \mathcal H=\frac14\left(\frac{A^2}{l^2}+\frac{F^2}{k^2}+\frac{2G^2}{k^2+l^2}\right).
\]
The above Nambu bracket is based on the Lie--Poisson bracket of $\mathfrak{so}(3)$, turning the Lorenz--1960 model into a particular form of the free rigid body equations. It therefore satisfies all properties of discrete Nambu mechanics. In this respect, the continuous Nambu bracket structure of the vorticity equation passes over to the discrete Nambu bracket structure of the free rigid body. Note, however, that the Lorenz--1960 model represents a restricted class of a free rigid body, as only two moments of inertia are independent. Moreover, it was demonstrated in~\cite{bihl09By} that the above truncation also respects the maximal set of admitted point symmetries in spectral space. This set of symmetries consists of discrete mirror transformations $(t,x,y,\psi)\mapsto (t,x,-y,-\psi)$ and $(t,x,y,\psi)\mapsto(t,-x,y,-\psi)$, together with combinations of shifts by $\pi$ in both $x$ and $y$ direction. These shifts are the admitted spectral counterparts of translational symmetries in physical space. In particular, using these transformations the above two observations by Lorenz can be naturally interpreted as conditions of symmetry. That is, for the Lorenz--1960 model preservation of symmetries and preservation of the Nambu structure are mutually compatible. Due to the Nambu representation, eqns.~\eqref{eq:Lorenz60} also satisfy Liouville's theorem.

As the Lorenz--1960 model inherits the Nambu structure of the vorticity equation and the selection of modes can be justified using the admitted point symmetries of the continuous equation, the notion of a maximum simplification may be regarded as appropriate.

\section{Structural properties of the Saltzman equations}\label{sec:RBC}

In this section, we discuss the structural properties of the convection model derived by Saltzman~\cite{salt62Ay}. That is, we discuss the admitted point symmetries and Nambu form and derive the maximum simplification that retains these properties in a minimal form.

The Saltzman equations we base our investigation on read in nondimensional form~\cite{hirs97Ay}:
\begin{equation} \label{eq:RBC}
	\pdl{\zeta}{t} + [\psi,\zeta] = R\sigma\pdl{T}{x} + \sigma\nn^4\psi, \qquad\qquad
    \pdl{T}{t} + [\psi,T]= \pdl{\psi}{x} + \nn^2 T.
\end{equation}
As before, $\psi$ is a stream function generating two-dimensional nondivergent flow in the $x$--$z$-plane, $\zeta = \nn^2\psi$ is the vorticity, $T$ is the temperature departure from a linear conduction profile, $\sigma$ is the Prandtl number, $R$ is the Rayleigh number and $[a,b] := \pds{a}{x}\,\pds{b}{z}-\pds{a}{z}\,\pds{b}{x}$ denotes the Jacobian.

In what follows, we aim to distinguish between dissipative and nondissipative systems. In the former, we preserve the form of equations as given in~\eqref{eq:RBC}, while in the latter we neglect terms $\nn^4\psi$, $\nn^2 T$. Note, however, that in the second case a different definition of $R$ and $\sigma$ arises, see~\cite{salt62Ay} for details.

Let us consider the domain $\Omega=[-L,L]\times[0,1]$. The boundary conditions we adopt are free-slip boundaries at both the top and the bottom of the fluid
\[
	\psi(t,x,z=0)=\psi(t,x,z=1)=0, \qquad \zeta(t,x,z=0)=\zeta(t,x,z=1)=0,
\]
together with
\[
	T(t,x,z=0)=T(t,x,z=1) = 0.
\]
Although it could be argued that non-slip boundaries in the vertical would be more natural to this viscous problem, the above choice is motivated to be able to incorporate the Lorenz--1963 model, which is based on free-slip boundaries. In $x$-direction there are different possibilities, e.g.\ periodic, free-stress or non-slip boundaries.

\subsection{Symmetries}

We are interested in point symmetries of system~\eqref{eq:RBC}. For this purpose, let us for the moment neglect the impact of the boundary conditions. To compute the maximal Lie invariance algebra we used the Maple package DESOLV \cite{carm00Ay}. The maximal Lie invariance algebra reads
\begin{subequations}\label{eq:symRBC}
\begin{align}
\begin{split}
  &\DDD = 2t\p_t + x\p_x + z\p_z - (3T-4Rz)\p_{T},\qquad  \p_{t},\qquad \p_{z}, \qquad \ZZ(g) = g(t)\p_{\psi},\\
	& \XX_1(f) = f(t)\p_{z} + f(t)R\p_{T} + f'(t)x\p_{\psi},\qquad \XX_2(h) = h(t)\p_{x} - h'(t)z\p_{\psi},
\end{split}	
\end{align}
where $f$, $g$, $h$ run through the set of real-valued time-dependent functions. Hence, system~\eqref{eq:RBC} admits scalings, shifts in $t$ and $z$, respectively, gauging of the stream function and generalized Galilean boosts in $z$- and $x$-direction, respectively. Moreover, there are two independent discrete symmetries given by
\begin{equation}
e_1\colon\ (t,x,z,\psi,T) \mapsto (t,x,-z,-\psi,-T), \quad e_2\colon\ (t,x,z,\psi,T) \mapsto (t,-x,z,-\psi,T).
\end{equation}
\end{subequations}
The presence of boundary conditions usually restricts the number of admitted symmetries strongly. In the symmetry analysis of differential equations, boundary value problems are rarely considered (see \cite{blum89Ay} for a discussion of this problem). On the other hand, for Rayleigh--B\'{e}nard convection, the consideration of boundaries obviously cannot be omitted. Therefore, we now single out those symmetries, which are admitted by the boundary value problem. We are only interested in symmetries acting on the space geometry of the problem. This is reasonable as transformations acting solely on $t$, $\psi$ or $T$ in the course of a series expansion do not place restrictions on the Fourier coefficients and thus cannot be used as a criterion for the selection of modes. This at once allows to exclude the transformations generated by $\p_t$ and $\ZZ(g)$ from our considerations. Moreover, scaling generated by $\DDD$ in any case would change the fixed geometry in $z$-direction, so we can exclude it too. The most general transformation generated by the remaining basis operators in combination with the discrete symmetries is given~by
\[
	(t,x,z,\psi,T)\mapsto (t,\delta_2(x+h\ve_2),\delta_1(z+f\ve_1+\ve_3),\delta_1\delta_2(\psi+f'x\ve_1-h'z\ve_2),\delta_1(T+Rf\ve_1)),
\]
where $\ve_i\in\mathbb{R}$ and $\delta_j\in\{-1,1\}$. Acting on the boundaries in $z$-direction, it is straightforward to determine those transformations preserving their values:
\begin{align}\label{eq:symboundary}
\begin{split}
	&(t,x,z,\psi,T) \mapsto (t,\delta_2(x+\ve_2),z,\delta_2\psi,T), \\
	&(t,x,z,\psi,T) \mapsto (t,\delta_2(x+\ve_2),1-z,-\delta_2\psi,-T).
\end{split}
\end{align}
It is now necessary to specify the boundaries in $x$-direction. A natural choice in the atmospheric sciences are periodic boundary conditions. This way, shifts in $x$-direction are admitted. On the other hand, this choice singles out the second discrete symmetry, i.e.\ we have $\delta_2=1$. This is the set of point symmetries on which we subsequently base our truncation.

\subsection{Nambu structure}

To make this paper self-contained, we restate some results given in~\cite{bihl08Ay}, slightly adapted for the special form of~\eqref{eq:RBC}. The conservative part of system~\eqref{eq:RBC} can be represented in continuous Nambu form by
\begin{subequations}
\begin{align}\label{eq:nambuRBCeq}
\begin{split}
   \pdl{\zeta}{t} &= -\left[\ddl{\mathcal{C}}{T},\ddl{\mathcal{H}}{\zeta}\right] - \left[\ddl{\mathcal{C}}{\zeta},\ddl{\mathcal{H}}{T}\right] = \{\zeta,\mathcal{C},\mathcal{H}\}, \\
   \pdl{T}{t} &= -\left[\ddl{\mathcal{C}}{\zeta},\ddl{\mathcal{H}}{\zeta}\right] \phantom{- \left[\ddl{\mathcal{C}}{\zeta},\ddl{\mathcal{H}}{T}\right]}\ = \{T,\mathcal{C},\mathcal{H}\},
\end{split}
\end{align}
with conserved quantities
\begin{equation}\label{eq:nambuRBCcons}
	 \mathcal{H} = \int_\Omega \left(\frac{1}{2}(\nabla\psi)^2 - R\sigma Tz\right)\ddd f, \quad \mathcal{C} = \int_\Omega\zeta(T-z)\,\ddd f,
\end{equation}
representing the total energy and a circulation-type quantity, respectively. In~\eqref{eq:nambuRBCeq}, the Nambu bracket $\{\cdot,\cdot,\cdot\}$ is defined for arbitrary functionals $\mathcal{F}_i = \mathcal{F}_i[\zeta,T], i=1,\dots, 3$ by the equation
\begin{align} \label{eq:nambuRBCbracket}
	\{\mathcal{F}_1,\mathcal{F}_2,\mathcal{F}_3\} := -\int_\Omega \Bigg( \ddl{\mathcal{F}_1}{T}\left[\ddl{\mathcal{F}_2}{\zeta},\ddl{\mathcal{F}_3}{\zeta}\right]+ \ddl{\mathcal{F}_1}{\zeta}\Bigg(\left[\ddl{\mathcal{F}_2}{T},\ddl{\mathcal{F}_3}{\zeta}\right] + \left[\ddl{\mathcal{F}_2}{\zeta},\ddl{\mathcal{F}_3}{T}\right]\Bigg)\Bigg) \ddd f.
\end{align}
This Nambu bracket is based on the semi-direct product extension of the Lie algebra of area-preserving diffeomorphisms with the vector space of real-valued functions on $\Omega$. The process of extension of a Lie--Poisson bracket is usually done for systems that incorporate more than one field variable. There are different ways how to extend a Lie algebra (see~\cite{thif00Ay} for an excellent overview), but the semi-direct extension is common for systems where one variable is advected by another (as is the temperature departure by the $\psi$-field). Using the Nambu bracket form, the semi-direct product structure can be cast in completely symmetric form.

We note that there is a second class of Casimir functionals given by
\begin{equation}\label{eq:nambuRBCcasimir}
	\mathcal{S}_g = \int_\Omega g\,\ddd f,
\end{equation}
where $g$ is an arbitrary function of $T-z$. This class of Casimirs is not needed for the conservative Nambu representation. However, it plays an important role for a geometric incorporation of dissipation. Namely, using the generalized free energy
\begin{equation}\label{eq:freeenergy}
	\mathcal{G}=\mathcal{H}-\mathcal{S} = \int_\Omega \left(\frac12(\nn\psi)^2-R\sigma Tz - \frac12R\sigma(T-z)^2\right)\ddd f,
\end{equation}
where $\mathcal{S}$ is a realization of the class of Casmirs $\mathcal{S}_g$, it is possible to represent the dissipative system~\eqref{eq:RBC} via
\begin{equation}\label{eq:nambuRBCdiss}
   \pdl{\zeta}{t} = \{\zeta,\mathcal{C},\mathcal{G}\}+\langle \zeta,\mathcal{G}\rangle,\quad   \pdl{T}{t} = \{T,\mathcal{C},\mathcal{G}\}+\langle T, \mathcal{G}\rangle,
\end{equation}
\end{subequations}
where
\[
	\langle \mathcal{F}_1,\mathcal{F}_2 \rangle := -\int_\Omega \left(\sigma\ddl{\mathcal{F}_1}{\zeta}\nn^4\ddl{\mathcal{F}_2}{\zeta} + \frac{1}{R\sigma}\ddl{\mathcal{F}_1}{T}\nn^2\ddl{\mathcal{F}_2}{T}\right)\ddd f
\]
is the symmetric bracket of dissipation, which is briefly discussed in the section below.

A delicate problem is to state appropriate boundary conditions making the Nambu bracket \eqref{eq:nambuRBCbracket} twofold antisymmetric. While the antisymmetry $\{\mathcal{F}_1,\mathcal{F}_2,\mathcal{F}_3\}= -\{\mathcal{F}_1,\mathcal{F}_3,\mathcal{F}_2\}$ is always satisfied due to the properties of the Jacobian operator, the antisymmetry $\{\mathcal{F}_1,\mathcal{F}_2,\mathcal{F}_3\}= -\{\mathcal{F}_2,\mathcal{F}_1,\mathcal{F}_3\}$ follows from an integration by parts. Hence, the boundary conditions must be specified in a way such that the resulting boundary terms vanish. For the specified boundary conditions in the vertical and periodical boundaries in $x$-direction, the Nambu bracket is indeed completely antisymmetric.

\subsection{Maximum simplification}

In this part, we derive the minimal model of~\eqref{eq:RBC} that retains both point symmetries and the associated Nambu form of the continuous equations. There are several attempts to extend the famous Lorenz--1963 model \cite{lore63Ay}. The motivation for these extensions is that the Lorenz--1963 model does not represent characteristic features of Rayleigh--B\'{e}nard convection properly, as noted e.g.\ in~\cite{chen06Ay}. Several authors have tried to improve the Lorenz--1963 model by attaching additional modes, but in various cases this lead to models exhibiting nonphysical behavior such as violation of energy or vorticity conservation (e.g.~\cite{howa86Ay}). The problem of energy conservation was solved in \cite{thif96Ay} where a universal criterion for the truncation to energy-conserving finite-mode models was established. Moreover, truncation to systems in coupled gyrostats form~\cite{gluh06Ay,gluh02Ay} may also lead to models that retain the conservation properties of the original equations. We also note, that a single gyrostat is a Nambu system and hence using such a truncation, conservation of the underlying geometry may be implemented at least in some minimal form.

It was shown in \cite{nevi94Ay} that the conservative part of the Lorenz--1963 model allows for a Nambu representation via
\[
	\dd{\mathbf{z}}{t} = \nabla_{\mathbf z}\mathcal{H}_1\times\nabla_{\mathbf z}\mathcal{H}_2,
\]
where $\mathbf{z} = (x,y,z)^\mathrm{T}$. The conserved quantities are
\[
	\mathcal{H}_1 = \frac12 x^2 - \sigma z, \quad \mathcal{H}_2 = \frac12 y^2 + \frac12 z^2 - rz,
\]
where $r=R/R_c$, with $R_c$ being the critical Rayleigh number. However, these two conserved quantities are proportional to spectral forms of energy and $\int_\Omega (T-z)^2\ddd f$, respectively, while the spectral expansion of $\mathcal{C}$ under the Lorenz ansatz gives identically zero. Therefore, the Lorenz--1963 truncation only allows for a Nambu form that is not directly related to the continuous Nambu form presented before. Moreover, if we also try to justify the selection of modes of the Lorenz--1963 system using point symmetries, we find that it would be necessary to simultaneously use the symmetries $e_2$ and shift in $x$-direction by $1$, which in any case would violate the boundary conditions. That is, the selection of modes is not natural from the symmetry point of view in this case.

Additionally, the Lorenz--1963 truncation does not account for the semi-direct product structure of the bracket of the continuous equations. An appropriate discrete realization of this semi-direct product structure is given by the special Euclidean algebra $\mathfrak{se}(3)=\mathfrak{so}(3)\ltimes\mathbb{R}^3$. The associated Lie--Poisson bracket on the dual $\mathfrak{se}(3)^*$ forms the basis of the Hamiltonian (or Nambu) representation of the heavy top equations in the body frame. The Lie--Poisson bracket reads~\cite{holm85Ay,mars99Ay}
\[
 \{F,G\} = -{\boldsymbol\Pi}\cdot (\nn_{\boldsymbol \Pi}F\times\nn_{\boldsymbol \Pi}G) - \boldsymbol{\Gamma}\cdot (\nn_{\boldsymbol \Pi}F\times\nn_{\boldsymbol \Gamma}G +\nn_{\boldsymbol \Gamma}F\times\nn_{\boldsymbol \Pi}G),
\]
where $\boldsymbol{\Pi}$ and $\boldsymbol{\Gamma}$ denote the vectors of angular momentum and the direction of gravity as seen from the body, respectively.

The heavy top model consists of three equations governing the evolution of angular momentum and three equations for the characterization of the direction of gravity as seen from the body. The maximum simplification of the Saltzman model based on the above Lie--Poisson bracket therefore needs a six-component reduction.

We now proceed with the construction of the modified Lorenz--1963 model. The expansion in Fourier series that is compatible with the specified boundary conditions is
\begin{align}\label{eq:expansion}
\begin{split}
	\psi &= \sum_{n=0}^{\infty}\sum_{m=1}^{\infty}\left(\phi_{nm}\sin an\pi x + \varphi_{nm}\cos a n \pi x\right) \sin m\pi z, \\
	T &= \sum_{n=0}^{\infty}\sum_{m=1}^{\infty}\left(\vartheta_{nm}\sin an\pi x + \theta_{nm}\cos a n \pi x\right) \sin m\pi z,
\end{split}
\end{align}
where $a$ is the inverse aspect ratio.

For the selection of modes for the six-component model, we employ the concept of symmetry in a similar fashion as in \cite{bihl09By,chen06Ay,hirs97Ay}. In these papers it was demonstrated that the admitted point symmetries of the original set of differential equations impose restricting conditions on the Fourier expansion that have to be taken into account in the course of the derivation of finite-mode models. As discussed above, the equations governing Rayleigh--B\'{e}nard convection admit an infinite-dimensional symmetry group with finite-dimensional subgroup~\eqref{eq:symboundary} preserving the boundary value problem. However, not all symmetry transformations included in~\eqref{eq:symboundary} may be used for a selection of modes since we cannot use all shifts in $x$-direction as the spectral space is essentially discrete.

For the selection of modes, we aim to use the transformations
\begin{align*}
\begin{split}
	t_1\colon\ & (t,x,z,\psi,T)\mapsto(t,x+1/a,z,\psi,T), \\
	t_2\colon\ & (t,x,z,\psi,T)\mapsto(t,x,1-z,-\psi,-T).
\end{split}
\end{align*}
The task is now to compute the corresponding implications of these transformations on the Fourier coefficients, which follow from a straightforward application to the expansion~\eqref{eq:expansion}. The transformation $t_1$ e.g.\ implies
\begin{align*}
  \psi &=  \sum_{n=0}^{\infty}\sum_{m=1}^{\infty}\left(\phi_{nm}\sin (an\pi x+n\pi) + \varphi_{nm}\cos (a n \pi x+n\pi) \right) \sin m\pi z,
\end{align*}
and similarly for the transformation of $T$. Hence $t_1$ leads to the spectral transformations $(\phi_{nm},\varphi_{nm},\vartheta_{nm},\theta_{nm})\mapsto ((-1)^n\phi_{nm},(-1)^n\varphi_{nm},(-1)^n\vartheta_{nm},(-1)^n\theta_{nm})$. In a similar fashion, the transformation $t_2$ is treated. The corresponding transformations in spectral space hence read
\begin{align}\label{eq:symsub}
\begin{split}
	t_1\colon\ &(\phi_{nm},\varphi_{nm},\vartheta_{nm},\theta_{nm})\mapsto ((-1)^n\phi_{nm},(-1)^n\varphi_{nm},(-1)^n\vartheta_{nm},(-1)^n\theta_{nm}), \\
	t_2\colon\ &(\phi_{nm},\varphi_{nm},\vartheta_{nm},\theta_{nm})\mapsto ((-1)^m\phi_{nm},(-1)^m\varphi_{nm},(-1)^m\vartheta_{nm},(-1)^m\theta_{nm}).
\end{split}
\end{align}
These two transformations give a restriction on the admitted modes since for all $n=2k-1$ and $m=2k-1$, respectively, the corresponding Fourier coefficients would violate the symmetry property and hence are not allowed in the truncation.

The discrete symmetry group generated by the transformations~\eqref{eq:symsub} is $G=\{e,t_1,t_2,t_1t_2\}$, where $e$ denotes the identity transformation. By exhaustively studying the implications of subgroups of $G$ on truncations of~\eqref{eq:expansion}, we can derive different low dimensional models in a similar fashion as was done in~\cite{bihl09By} for the barotropic vorticity equation. The list of nontrivial subgroups of $G$ is given by $S_1=\{e,t_1\}$, $S_2=\{e,t_2\}$ and $S_3=\{e,t_2t_3\}$.

Since we already know that the model to be derived must have six coefficients, it remains to select them in accordance with the above subgroups. The first six nonvanishing coefficients under consideration of the subgroup $S_1$ are $\varphi_{01}$, $\theta_{01}$, $\phi_{21}$, $\varphi_{21}$, $\vartheta_{21}$ and $\theta_{21}$. Since these coefficients do not include those of the original Lorenz--1963 model, the resulting model will not be considered here. Selecting the modes using the subgroup $S_2$ and $G$ itself also does not allow to incorporate the Lorenz--1963 model.

The remaining possibility is given by the subgroup $S_3$, leading to the choice of coefficients $\phi_{11}$, $\varphi_{11}$, $\vartheta_{11}$, $\theta_{11}$, $\varphi_{02}$, $\theta_{02}$. This choice incorporates both the Lorenz--1963 model and the model in~\cite{chen06Ay} and also gives a sound justification for the selection of modes. In addition to the symmetries, we also aim to preserve the semi-direct product structure of the Nambu representation~\eqref{eq:nambuRBCbracket}. For this purpose, the selection of the above listed coefficients based on symmetry considerations is still too general. It is necessary to scale these coefficients appropriately. Setting
\begin{align*}
	&\phi_{11} = bA,&  &\varphi_{11} = bB,&  &\varphi_{02} = cC,& \\
	&\vartheta_{11} = eD,& &\theta_{11} = eE,&  &\theta_{02} = fF,
\end{align*}
and plugging the corresponding truncation of~\eqref{eq:expansion} into the conservative part of system~\eqref{eq:RBC}, it is found that the scaling coefficients have to satisfy
\[
	c = \frac{1}{2b}, \quad e = \frac{a^3}{\pi^2(1+a^2)},\quad f = \frac{2a^3}{\pi^2b^2(1+a^2)^2},
\]
in order to allow for a Nambu representation of heavy top form:
\[
	\{\mathcal F_1,\mathcal F_2, \mathcal F_3\} := -\nn_{\boldsymbol{\Gamma}}\mathcal F_1 \cdot \nn_{\boldsymbol{\pi}}\mathcal F_2 \times\nn_{\boldsymbol{\pi}}\mathcal F_3-\nn_{\boldsymbol{\pi}}\mathcal F_1 \cdot (\nn_{\boldsymbol{\Gamma}}\mathcal F_2\times\nn_{\boldsymbol{\pi}}\mathcal F_3+ \nn_{\boldsymbol{\pi}}\mathcal F_2\times\nn_{\boldsymbol{\Gamma}}\mathcal F_3).
\]
Here, ${\boldsymbol \pi} = (A,B,C)^{\mathrm T}$ and ${\boldsymbol \Gamma} = (D,E,F)^{\mathrm T}$ are the fluid mechanical analogs of the vector of angular momentum and the direction of gravity as seen from the body, respectively.

The resulting six-component model then reads
\begin{align}\label{eq:lagrangetop}
\begin{split}
	\dd{A}{t} &= \frac{a}{2b\pi(1+a^2)}((a^2-3)\pi^3 BC+2eR\sigma E), \\
	\dd{B}{t} &= -\frac{a}{2b\pi(1+a^2)}((a^2-3)\pi^3 AC+2eR\sigma D), \\
	\dd{C}{t} &= 0, \\
	\dd{D}{t} &= \frac{a\pi}{2be}( e\pi CE-2 b^2 f\pi BF- 2b^2B), \\
	\dd{E}{t} &= -\frac{a\pi}{2be}(e\pi  CD - 2 b^2 f\pi A F - 2b^2A), \\
	\dd{F}{t} &= \frac{abe\pi^2}{2f}(BD-AE).
\end{split}
\end{align}
The conserved quantities~\eqref{eq:nambuRBCcons} are correspondingly
\begin{align*}
	&\mathcal H = \frac{1}{4ab^2\pi}((1+a^2)b^4\pi^3(A^2+B^2)+2\pi^3C^2+4Rb^2f\sigma F), \\
	&\mathcal C = -\frac{\pi}{2ab}((1+a^2)b^2e\pi(AD+ BE)+4f\pi CF+4C).
\end{align*}
Using both the heavy top Nambu bracket and the conserved quantities, the finite-mode model~\eqref{eq:lagrangetop} can be cast in Nambu from
\begin{align*}
   \dd{\mathbf x}{t} = \{\mathbf x,\mathcal C,\mathcal H\},
\end{align*}
where $\mathbf{x} = (A,B,C,D,E,F)^{\mathrm T}$. Note, that this Nambu bracket formulation is now structurally completely analog to the continuous Nambu bracket~\eqref{eq:nambuRBCbracket} of the convection equations, as it is based on the Lie--Poisson bracket using the Lie algebra $\mathfrak{se}(3)$. In particular, by fixing one argument in the heavy top Nambu bracket one naturally recovers all the Hamiltonian properties of the Lie--Poisson system, e.g.\ the Jacobi identity. Bracket~\eqref{eq:nambuRBCbracket} therefore also automatically conserves the corresponding truncated forms of the second class of Casimirs~\eqref{eq:nambuRBCcasimir}, such as
\[
 \mathcal S = \frac{R\sigma}{12a\pi}\left(3e^2\pi (D^2+E^2)+6f^2\pi F^2+12fF\right).
\]
It is remarkable that the above model has now a mechanical interpretation similar to the Lorenz--1960 model. As the Lorenz--1960 model is a restricted class of the free rigid body equations, the finite-mode model~\eqref{eq:lagrangetop} in turn may be considered as a restricted class of the heavy top equations that is referred to as \textit{Lagrange top} \cite{holm85Ay}. This is a heavy top with two moments of inertia being equal (a symmetric top) with the position vector of the center-of-mass pointing in $C$-direction (which leads to $\ddd C / \ddd t = 0$). Correspondingly, a number of results valid for the Lagrange top might already be passed over to the model~\eqref{eq:lagrangetop}, such as issues of stability~\cite{holm85Ay} or numerical algorithms preserving the Lie--Poisson structure of the above bracket~\cite{engo02Ay}. Moreover, due to the additional conserved quantity $C=\const$ the Lagrange top is a prominent example of a Liouville integrable system~\cite{audi99Ay,gavr98Ay}. In addition, since $\pds{\dot x_i}{x_i}=0$, $\forall i=1,\dots,6$, the above set of equations also satisfies the Liouville theorem.

Note that the term linear in $C$ (resp.\ $F$) in the expression for $\mathcal C$ (resp.\ $\mathcal S$) arises due to the use of variable $T$ instead of $\tilde T = T-z$ and correspondingly do the terms linear in $A$ and $B$ in the fourth and fifth equation, respectively.

If $C=0$ at the onset of evolution, the above model reduces to the five-component model given in \cite{chen06Ay} upon rescaling of the Fourier modes. Although in any case $C=\const$ during evolution, we find it nevertheless important to retain this component in the above model. Firstly, it enables to cast the reduced model in Lagrange top form. Secondly, the selection of modes based on symmetries does not allow to truncate the Saltzman equations to a five component model, since there is no additional criterion that permits one to predict a priori whether to chose $\varphi_{02}$ or $\theta_{02}$. Hence, both coefficients must be incorporated in the Fourier series expansion.

The extension to a Nambu-metriplectic finite-mode model is straightforward. Following \cite{morr86Ay}, a discrete metric system can defined via $\ddd \textbf{z}/\ddd t = g \nabla_{\mathbf z} P=\left\langle \textbf{z}, P\right\rangle$, where $\textbf{z}$ is the phase space vector, $P$ denotes a phase space function and $g$ is a tensor. It is further required that $\left\langle F_1, F_2 \right\rangle=\left\langle F_2, F_1 \right\rangle, \forall F_1, F_2$, which in turn enforces $g$ to be symmetric. In our case, using the above truncation and correspondingly the generalized free energy~\eqref{eq:freeenergy} as phase space functional, we can incorporate the dissipative terms upon using
\[
	g =-2a\,\mathrm{diag}\left(\frac{\sigma}{b^2},\frac{\sigma}{b^2},2\sigma b^2,\frac{\pi^2(1+a^2)}{Re^2\sigma},\frac{\pi^2(1+a^2)}{Re^2\sigma}, \frac{2\pi^2}{Rf^2\sigma}\right)
\]
as metric tensor. Then, attaching $g\nabla_{\mathbf x}\mathcal G$ to system~\eqref{eq:lagrangetop} gives the maximum simplification of the dissipative Saltzman equations, which reads
\begin{align}\label{eq:lagrangetopdiss}
\begin{split}
	\dd{A}{t} &= \frac{a}{2b\pi(1+a^2)}((a^2-3)\pi^3 BC+2eR\sigma E)-(1+a^2)\pi^2\sigma A, \\
	\dd{B}{t} &= -\frac{a}{2b\pi(1+a^2)}((a^2-3)\pi^3 AC+2eR\sigma D)-(1+a^2)\pi^2\sigma B, \\
	\dd{C}{t} &= -4\pi^2\sigma C, \\
	\dd{D}{t} &= \frac{a\pi}{2be}( e\pi CE-2 b^2 f\pi BF- 2b^2B)-(1+a^2)\pi^2 D, \\
	\dd{E}{t} &= -\frac{a\pi}{2be}(e\pi CD-2 b^2 f\pi AF- 2b^2A)-(1+a^2)\pi^2 E, \\
	\dd{F}{t} &= \frac{abe\pi^2}{2f}(BD-AE)-4\pi^2 F,
\end{split}
\end{align}
or in the more compact Nambu-metriplectic form
\[
    \dd{\mathbf x}{t} = \{\mathbf x,\mathcal C,\mathcal G\}+\langle \mathbf x,\mathcal G\rangle.
\]
System~\eqref{eq:lagrangetopdiss} can be considered as a damped Lagrange top and its Nambu-metriplectic form completes the geometric picture of the maximal structure-preserving truncation of the dissipative Saltzman equations~\eqref{eq:nambuRBCdiss} discussed in the present paper.

It was noted in~\cite{herm95Ay} that any vorticity field under the Boussinesq approximation has to satisfy the balance equation:
\[
	\pdl{}{t}\int\limits_0^1\int\limits_0^{2/a}\zeta\, \ddd x \ddd z = \sigma \Bigg[\pdl{}{z}\int\limits_0^{2/a}\zeta\ddd x\Bigg]\Bigg|_{z=0}^{z=1}.
\]
Straightforward computation for the six-component model shows that this balance equation is identically satisfied for any value of the Prandtl number.

It should be emphasized that the results in this section were derived under the assumption of $\delta_2=1$ in~\eqref{eq:symboundary}. This choice was enforced due to the use of periodic boundary conditions, which are of obvious importance in geophysical fluid dynamics. On the other hand, the alternative choice of $\delta_2=-1$ would not allow to derive the above discrete convection model, as it would require a different selection of Fourier modes (not including the Lorenz--1963 model). However, this is quite natural as a physical realization admitting this reflection symmetry involves sidewalls at $x=-L,L$~\cite{hirs97Ay}. For such a configuration, the generic Nambu structure~\eqref{eq:nambuRBCbracket} has to be supplemented with boundary term contributions, since the second antisymmetry relying on integration by parts is then not automatically fulfilled any more. As usual in continuous Hamiltonian and Nambu mechanics, the presence of nontrivial boundary conditions complicates the appropriate formulation of the models, which in the present case also passes over to the finite-mode simplification.

\section{Conclusion and outlook}\label{sec:conclusion}

In this paper we have addressed the problem of maximum simplification of atmospheric models by a discussion of the Lorenz--1960 and Lorenz--1963 model. It was reviewed that the Lorenz--1960 model is indeed the maximum simplification of the inviscid barotropic vorticity equation that preserves both point symmetries as well as the inherited Nambu structure of the continuous counterpart. This way, we have also implicitly shown that the selection of modes in accordance with symmetries is compatible with the inherited Nambu form of the discrete model. Inspired from this tutorial example, the Lorenz--1963 model was investigated too. It was found that this model neither preserves the proper Nambu structure nor is it compatible with respect to the underlying symmetries. This may serve as an additional justification of reported unphysical behavior of this model. The proposed extension of the Lorenz--1963 model is a six-component truncation that also includes the model~\cite{chen06Ay} as special case. This model is constructed using a subgroup of the symmetry group of the Saltzman equations preserving the boundary value problem. Moreover, the semi-direct product structure of the Lie--Poisson (or Nambu) bracket of the field equations is retained, hence the model is automatically energy- and Casimir-conserving in the nondissipative limit. This again implicitly shows that for the presented six-component truncation both the Nambu structure and the admitted point symmetries are compatible. Incorporation of dissipation leads to a discrete Nambu-metriplectic model, also conserving the symmetric structure given by the metric part of the continuous bracket. This compatibility of the six-component model with geometric structures may be considered beneficial in view of the testing issues reviewed in section~\ref{sec:introduction}, for which the newly derived model could be employed. Moreover, due to the preservation of important geometric structures, both models that were presented in this paper may deserve the notion of a maximum simplification.

If one aims to use finite-mode models for physical purposes and not only as toy models, the question of structure-preserving extensions of such minimal systems of equations is of certain interest. It was indicated in section~\ref{sec:nambu} that there is merely the method of Zeitlin that allows to construct fully Hamiltonian finite-mode approximations of 2d fluid mechanics. In this method, series of approximated $n$-dimensional models are derived on the two-dimensional torus $T^2$, which in the limit $n\to\infty$ converge to the vorticity equation and the equations of a Boussinesq stratified fluid, respectively~\cite{zeit91Ay,zeit05Ay}. This convergent sequence of finite-dimensional models exists due to the property that for a certain representation of $SU(n)$, in the limit of $n\to\infty$, the group of area-preserving diffeomorphisms (and its semi-direct extension by a vector space, respectively) is recovered. One main benefit of this method is that for each finite-mode model a maximal number of Casimirs is preserved, which makes such models very attractive, e.g.\ for the investigation of statistical properties of fluid mechanical systems~\cite{abra03Ay}. For more general settings than $T^2$ or for three-dimensional fluid mechanics, however, the question of a connection between finite- and infinite-dimensional Lie--Poisson systems is not fully answered yet. This points to another still unsolved question, namely whether it is possible to relate discrete and continuous Nambu mechanics in some natural way. Although such a relation was established for two very low dimensional models in the present paper, the problem of a proper extension of these minimal models without violating the Nambu structure has not been tackled yet. Furthermore, it would be interesting to apply the method used in this paper to other models of fluid mechanics. This way, a list of maximal simplified structure-preserving models could be established. This should be the issue of forthcoming work.

\section*{Acknowledgments}

The authors acknowledge the useful discussions with K.\ Brazda, M.\ Hantel, P.\ N\'{e}vir, R.O.\ Popovych and M.\ Sommer. AB is a recipient of a DOC-fellowship of the Austrian Academy of Sciences. The research of JS was supported by the Austrian Science Fund (FWF), project P21335. 

\end{document}